\documentclass[12pt,a4j,notitlepage,fleqn]{article}

\usepackage{setspace}
\usepackage[margin=1in]{geometry}
\usepackage[symbol]{footmisc}
\usepackage[T1]{fontenc}
\usepackage[utf8]{inputenc}
\usepackage{authblk}

\makeatletter

\newcommand{\fboxsubsec}[1]{
	\begin{flushleft}
		#1
	\end{flushleft}
	}
\newcommand{\fboxsubsubsec}[1]{
	\begin{flushleft}
		#1
	\end{flushleft}
	}
\renewcommand{\subsection}{\@startsection{subsection}{2}{0pt}
	{1ex}
	{0.5ex}
	{\reset@font\it\fboxsubsec}
	}
\renewcommand{\subsubsection}{\@startsection{subsubsection}{2}{0pt}
	{1ex}
	{0.5ex}
	{\reset@font\fboxsubsubsec}
	}
\makeatother

\title{Market Efficiency and Government Interventions\\
in Prewar Japanese Rice Futures Markets}%

\author{Mikio Ito$^{a}$, \ Kiyotaka Maeda$^{b}$ \ and \ Akihiko Noda$^{c}$\thanks{\scriptsize Corresponding Author. E-mail: noda@cc.kyoto-su.ac.jp, Tel. +81-75-705-1510, Fax. +81-75-705-3227.}

{\scriptsize ${}^{a}$ \it Faculty of Economics, Keio University, 2-15-45 Mita, Minato-ku, Tokyo 108-8345, Japan}

{\scriptsize ${}^{b}$ \it Faculty of Economics, Seinan Gakuin University, 6-2-92 Nishijin, Sawara-ku Fukuoka 814-8511, Japan}

{\scriptsize ${}^{c}$ \it Faculty of Economics, Kyoto Sangyo University, Motoyama, Kamigamo, Kita-ku, Kyoto 603-8555, Japan}

{\scriptsize ${}^{d}$ \it Keio Economic Observatory, Keio University, 2-15-45 Mita, Minato-ku, Tokyo 108-8345, Japan}}
\date{This Version: \today}

\renewcommand\thefootnote{\arabic{footnote}}

\usepackage[dvips]{graphicx}

\usepackage[]{natbib}%
\usepackage{amsmath,amssymb}%
\usepackage{ascmac}%
\usepackage{multirow}%
\usepackage{lscape}%
\usepackage{subfigmat}

\usepackage{pifont}%
\usepackage{arydshln}%
\usepackage[format=hang]{caption}
\usepackage[all]{xy}
\usepackage{url}
\bibpunct{(}{)}{;}{a}{}{,}

\def\hsymbu#1{\smash{\lower1.7ex\hbox{\huge$#1$}}}

\def\ve #1{{\mbox{\boldmath $#1$}}}

\newcommand{\citetapos}[1]{\citeauthor{#1}'s \citeyearpar{#1}}
\newcommand{\citeapos}[2]{\citeauthor{#1}'s (\citeyear{#2})}

\def\ve #1{{\mbox{\boldmath $#1$}}}

\begin{document}

\begin{titlepage}

\renewcommand{\thepage}{}
\renewcommand{\thefootnote}{\fnsymbol{footnote}}

\maketitle

\vspace{-10mm}

\noindent
\hrulefill

\noindent
{\bfseries Abstract:} This study analyzes how colonial rice trade in prewar Japan affected its rice market, considering several government interventions in the two rice futures exchanges in Tokyo and Osaka. We explore the interventions in the futures markets using two procedures. First, we measure the joint degree of efficiency in the markets using a time-varying vector autoregression model. Second, we examine many historical events that possibly affected the markets and focus on one event at a time. The degree varies over time within our sample period (1881--1932). The observation, together with historical analysis, leads to the following conclusions. (1) The two major markets in Tokyo and Osaka were nearly efficient. (2) Government interventions involving the delivery of imported rice from Taiwan and Korea often reduced futures market efficiency. Finally, (3) this relationship continued as long as the quality difference between imported and domestic rice existed. The government interventions that promoted domestic distributions of the colonial goods resulted in confusion in the commodity markets, and decreased efficiency of the markets in the metropole.\\

\noindent
{\bfseries Keywords:} Futures Market; Colonial Trade; Government Intervention; Market Efficiency.\\

\noindent
{\bfseries JEL Classification Numbers:} N25; G13; G14.

\noindent
\hrulefill

\end{titlepage}


\bibliographystyle{asa}%

\pagebreak

\section{Introduction}\label{ei_sec1}

Economic historians have long been interested in how colonial trade of advanced countries affected markets in metropoles. From the 17th century, some European countries colonized countries in the Americas, Africa, and Asia. These colonizing countries procured food and resources from their colonies, and colonial trade expanded. In the 19th century, colonial trade began to help advanced countries to save specie money under the gold standard. Accordingly, these countries promoted imports from their colonies by changing tariff policies. At the same time, their commodity markets improved.

Steam locomotives and steamships were developed significantly in the 19th century. These new transportation modes made it possible to carry cargo between long distances in bulk and rapidly. As a result, commercial transactions, including foreign trade, grew. However, the growing transactions allowed traders to face increasing risk of price volatility. Accordingly, futures exchanges were organized in these countries. For example, the Chicago Board of Trade was founded in 1848 (see \citet[p.11]{kaufmann1984hfm}). The futures exchanges provided index prices for the corresponding spot markets, and contributed to the development of commodity markets. Many previous studies have paid attention to futures trading on the commodity exchanges.

The principal focus of this body of literature has been to test the efficient market hypothesis. Through such tests, scholars have examined the performance of risk hedging in modern and contemporary futures markets for agricultural commodities, livestock, metals, oil, gas, and foreign exchange. In particular, \citet{perkins1974efm} and \citet{jacks2007pvt} consider the modern futures markets for agricultural commodities in Osaka and Chicago using data on futures prices. However, there are very few studies that investigate how growing colonial trade affected the efficiency of commodity markets in home countries. After the 19th century, the development of commodity markets was promoted not only by the introduction of the new transportation modes but also by increased imports of commodities from the colonies. Nevertheless, much of the literature focuses on whether the efficient market hypothesis holds for a given observation. Thus, it cannot discuss the relationship between growing colonial trade and continual change in the commodity markets over time. Accordingly, this study focuses on the rice market in Japan before World War {I\hspace{-.1em}I} (WW{I\hspace{-.1em}I}), (henceforth, ``prewar Japan'') for three reasons, as follows.

First, goods supply of commodity markets in the home islands of Japan strongly depended on its colonies (see \citet{okubo2007tbf} and \citet[pp.72--75]{hori2009hce}). Second, the Japanese government accelerated rice imports from its colonies, Taiwan and Korea, because rice has been the staple food of Japan since ancient times. However, since industrialization and urbanization were advanced in Japan in the 1890s, rice supply had been insufficient in Japan. Consequently, the Japanese government promoted rice imports from Taiwan and Korea to satisfy the shortage. Third, since the early 18th century, merchants had traded rice in futures actively. In particular, \citet{schaede1989fft} and \citet{blank1991fom} describe the earliest Japanese futures exchange, Dojima Kome Kaisho (i.e., the Osaka–Dojima rice exchange), as one of the oldest examples of a well-organized futures market. The exchange developed with the official approval of the Tokugawa Shogunate, which sometimes intervened in the exchange to control rice prices (see \citet[ch.3]{takatsuki2008crt}). Succeeding the Tokugawa Shogunate, the Meiji government strengthened the political function of the Japanese capital, Tokyo, and in 1871, another rice futures exchange was established there by rice brokers and major merchants. This began a period in which there were two major rice exchanges, in Tokyo and Osaka. The government was always concerned about the rice market and often intervened in the exchanges to suppress rice prices.

Before WW{I\hspace{-.1em}I} in Japan, variation in rice prices was the main cause of the variation in general price levels. In fact, the rice price accounted for 13 per cent of the weight of the Tokyo Wholesale Price Index in the 1933 Base. This was twice as high as the weight of cotton yarn, the second largest weight, which was a major export item in Japan (see \citet[p.40]{boj1987hys}). Accordingly, as we discuss in Section \ref{ei_sec2}, government intervention to suppress rice prices had great significance for not only food supply policy but also price stabilization policy. Under the circumstances, the government ordered the exchanges to change their transaction rules regarding rice imported from Japanese colonies. However, previous studies have paid little attention to the relationship between the government-led amendment of the transaction rules and the continual change in commodity markets (see \citet[pp.125--145]{taketoshi1999ear} and \citet[pp.25--45]{shizume2011fep}). 

Regarding this amendment as a kind of government intervention, we address policy changes that would affect the efficiency of the rice futures exchanges, which were well-organized markets, in prewar Japan. To such a setting, one might apply a test of semi-strong efficiency in the line of \citet{fama1970ecm}, considering policy changes to be added into an information set of market participants. Since \citet{fama1970ecm} founded the research program for the efficiency of financial markets, a vast body of literature has developed, including studies analyzing the semi-strong efficiency of financial markets, such as spot and futures markets for foreign exchange. For instance, following \citetapos{fama1970ecm} categorization of market efficiency, \citet{hansen1980fer} formalize a method for testing both weak and semi-strong forms of efficiency in foreign exchange markets. Some studies, in the context of a semi-strong form of efficiency, examine whether the rate of return is affected by some factors other than the lagged rates. In practice, \citet{frenkel1980erp} studies the efficiency of foreign exchange markets under floating systems, including semi-strong settings. 
\citet{goss1987wpp} examines whether Australian wool futures prices reflect available information by regressing the forecast error to other markets' forecast errors. 
\citet{colling1990rlh} study how live hog futures prices react to anticipated and unanticipated information of the prices by using two-limit tobit models.

However, the abovementioned approaches are based on the semi-strong form of efficiency and are unsuitable for our case. Suppose that a certain policy affecting financial markets can be detected by using statistical tests for the semi-strong form, as reviewed above. Two cases exist. First and foremost, numerical time-series data for the policy variable are known and available for its regression analysis. Second, changes of regulations and rules, such as those we address in this study, occur at a known period so that we can form dummy variables for each period to conduct regression analysis. Both cases are troublesome in our case, since both effective policies and their effective periods are hardly known without examining historical events that possibly affected the rice futures markets in prewar Japan.

Thus, this study's analysis concentrates exclusively on market efficiency in the weak sense. This study makes a significant contribution to economic history by featuring time-varying efficiency in the weak sense. Here, we present our reasons for choosing this methodology. First, the government interventions addressed in this study involve changes in traders' minds through a behavioral manner. According to \citeapos{lo2004amh}{lo2004amh,lo2005rem} new concept in the field of behavioral finance, known as adaptive market hypothesis (AMH), we consider that possible variations in market efficiency reflects such mind changes. The AMH states that traders in financial markets change their minds not through a deductive manner but through a behavioral manner. The former is based on rational expectations and the latter reflects psychological and seemingly irrational factors. Replacing the efficient market hypothesis (EMH) with his AMH, Lo and his followers shed new light on real financial markets that are supposedly efficient at one time and inefficient at another time (see for example \citet{lo2012amn}, \citet{neely2009amh} and \citet{noda2016amh}). Our methodology has some points in common with his approach based on the AMH. In practice, we consider that varying efficiency of rice markets in prewar Japan reflected changes in traders' minds that were caused by government interventions. At the same time, the government interventions in the rice markets themselves were not always understandable from both economic and financial viewpoints. Second, although the semi-strong form approach has been significant for an inquiry into market efficiency, recent literature about market efficiency shows us that few recent researchers study market efficiency in the semi-strong form context for the following possible reason (see a recent survey paper by \citet{lim2011esm}, for instance).

We explore qualitative policy changes in the futures markets using the following two procedures. First, we examine the efficiency of the markets varying over time using the time-varying VAR model. Second, we examine many historical events that possibly affected the markets and focus on one event at a time. It should be stressed that without the first procedure, the list of such events and periods cannot be narrowed down.

In practice, after presenting a short history of the rice futures market in prewar Japan, we measure the time-varying degree of market efficiency in rice futures markets using \citetapos{ito2014ism} non-Bayesian time-varying vector autoregression (TV-VAR) model. The degree indicates the relative efficiency of the rice futures market in each period, in the weak sense of \citet{fama1970ecm}, when we regard the market as being highly financial. Using inference based on applying the bootstrap technique to the resulting degree estimates, we detect periods in prewar Japan in which the rice futures market was relatively inefficient. In particular, in 1890, 1898, several years after 1912, and from 1918 to 1921, less efficient markets were observed. We explore government interventions in the rice futures exchanges in Tokyo and Osaka that correspond to the periods using historical documents. These interventions were rooted in the government's concern about imported rice from Korea and Taiwan. Then, the government forced the exchanges to accept imported rice from Taiwan and Korea, which was of lower quality than domestic rice. This resulted in reduced market efficiency in the two major rice futures exchanges, since traders were upset by the changes in the trading rules.

The rest of this paper is organized as follows. Section \ref{ei_sec2} introduces some features of the two major rice futures markets in the prewar period and summarizes key facts of Japanese history associated with these markets. In Section \ref{ei_sec3}, we present \citetapos{ito2014ism} non-Bayesian TV-VAR model. Section \ref{ei_sec4} describes the data, covering the two major rice futures markets in prewar Japan, and presents preliminary unit root test results. Section \ref{ei_sec5} presents the empirical results using time-invariant and time-varying VAR models. In Section \ref{ei_sec6}, we discuss market efficiency in these exchanges and argue that some government interventions disrupted the futures markets and reduced their efficiency. Section \ref{ei_sec7} concludes.

\section{A History of Rice Futures Markets in Japan}\label{ei_sec2}

In prewar Japan, two major rice futures exchanges, the Tokyo rice exchange and the Osaka-Dojima rice exchange, were considered leading markets. Rice trading volumes in the two exchanges amounted to 18.38 million {\it koku}, or 45 per cent of total rice futures, in 1893.\footnote{{\it Koku} is a unit of rice trading volume. One {\it koku} is equal to 180.39 liters.} The two major exchanges continued to rank first and second in rice futures trading volumes (see \citet[pp.154--159]{mac1900sya}). They listed each single standard brand of domestic rice: Musashi (rice cropped in Saitama prefecture) on the Tokyo rice exchange and Settsu (rice cropped in Hyogo prefecture) on the Osaka-Dojima rice exchange.\footnote{The Saitama prefecture is next to the Tokyo prefecture and the Hyogo prefecture is next to the Osaka prefecture.}

These standard brands were regarded as the basis of transactions, and both brands were equal in quality
 (see \citet[p.457]{ichinoue1920ssr}). In fact, the Tokyo rice exchange categorized Settsu rice in the same groups as Musashi rice in 1920 (see \citet[pp.458-459]{ichinoue1920ssr}). These exchanges approved two manners for clearing rice futures transactions: settlement on the balance and delivery of physical rice. It is an interesting feature of rice futures transactions in Japan that sellers often delivered physical rice to buyers when they cleared their transactions in futures. In fact, the average ratios of delivery volumes to trading volumes from 1912 to 1931 were six per cent for the Tokyo rice exchange and five per cent for the Osaka-Dojima rice exchange. In the 1910s, these ratios were higher than after the 1920s. From 1912 to 1919, these ratios were nine per cent in Tokyo and eight per cent in Osaka (see \citet{maf1935tvp2,maf1935tvp4,maf1935tvp1}).

The exchanges set the standard quantities of rice delivery for 100 {\it koku}, and the sellers had to complete the delivery by the end of each month. Specifically, the sellers put the delivery rice in warehouses that were designated by the exchanges and located in the same cities as the exchanges. The warehouses issued warehouse receipts to sellers, and the sellers handed the receipts to buyers in the exchanges (see \citet{doujima1934or} and \citet{tokyo1934or}). In principle, the exchanges permitted traders to use only domestic rice for deliveries. Each November, the exchanges inspected the quality of domestic rice, and captured the quality differences between standard rice and any other domestic rice type. Based on the results of the quality inspection, the exchanges provided a correspondence table for rice traders who delivered physical rice of different grades and from different production areas. This classification was provided for actual rice delivery among traders. The exchanges classified the delivery rice by the following two criteria: place of origin and quality. For example, in 1920, the Tokyo rice exchange separated delivery rice into eight groups based on place of origin. Furthermore, the exchange grouped delivery rice into 16–19 classes by place of origin as well as quality. As a result, the Tokyo rice exchange classified delivery rice into 165 groups in 1920 (see \citet[pp.455-458]{ichinoue1920ssr}). When traders, sellers, or buyers delivered rice that was different in quality from the standard, they had to pay an amount of cash corresponding to the difference in grade. In particular, when sellers delivered rice inferior to the standard rice, sellers had to pay buyers a balance according to the grade difference. By contrast, when sellers delivered non-standard rice of superior grade, buyers had to pay the sellers an amount of cash that was subject to the difference in grade. Thus, the sellers could deliver non-standard rice of superior and inferior grade. However, the buyers could not recognize which rice would be delivered until they received the warehouse receipts in the exchanges. This method of trading--that is, trading by grade--was revised along with changes in the rice market.

In the late 1880s, rice exports from Japan were expanded. From 1885 to 1889, the amount of rice exported increased from 0.1 million {\it koku} to 1.3 million {\it koku}. This growing export volume resulted from an upturn in the terms of trade, caused by stable silver prices, and a decrease in the price of domestic rice at the time (see \citet[pp.18--20]{omameuda1993fpm}). However, the balance between rice supply and demand was lost in Japan after the same period.

From the late 1880s to the late 1890s, the volume of rice production increased by 7 per cent while the total population in Japan grew by 17 per cent. In particular, the urban population increased rapidly. From the same period, the Japanese government and major railway companies expanded transportation infrastructure, such as roads, railways, and port facilities. This infrastructure accelerated the concentration of population and distribution in the major cities. For example, from 1886 to 1898, the population of the Tokyo city increased by 28 per cent and that of Osaka city by 27 per cent (see \citet[pp.507, 634, 642--645]{toyo1927jss}).\footnote{We provide the detailed data of the rice production and the population in Japan from the late 1880s to the late 1890s. The five years average of rice production volume in the late 1880s (1885-89) and the late 1890s (1895-99) were 36,577,286 {\it koku} and 39,265,273 {\it koku}, respectively. The number of total population in 1885 and 1899 were 37,868,949 and 44,270,495, respectively. The population in the city of Tokyo were 1,121,883 in 1886 and 1,440,121 in 1898; those in the city of Osaka were 361,694 in 1886 and 821,235 in 1898.} In short, both Tokyo and Osaka had rapidly growing populations, and these cities had become Japanese centers of distribution and consumption. As for rice distribution, Tokyo and Osaka were the two centers of rice circulation in Eastern Japan and Western Japan, respectively (see \citet[pp.122--128]{kataoka2004pdc}). The increase in population, led by urban growth, affected the balance of rice supply and demand, and rice exports from Japan shrank at the end of the 1880s.

\begin{figure}[bp]
 \caption{Spot Prices in Tokyo and Osaka Rice Markets}\label{ei_fig1}
 \begin{center}
 \includegraphics[scale=0.7]{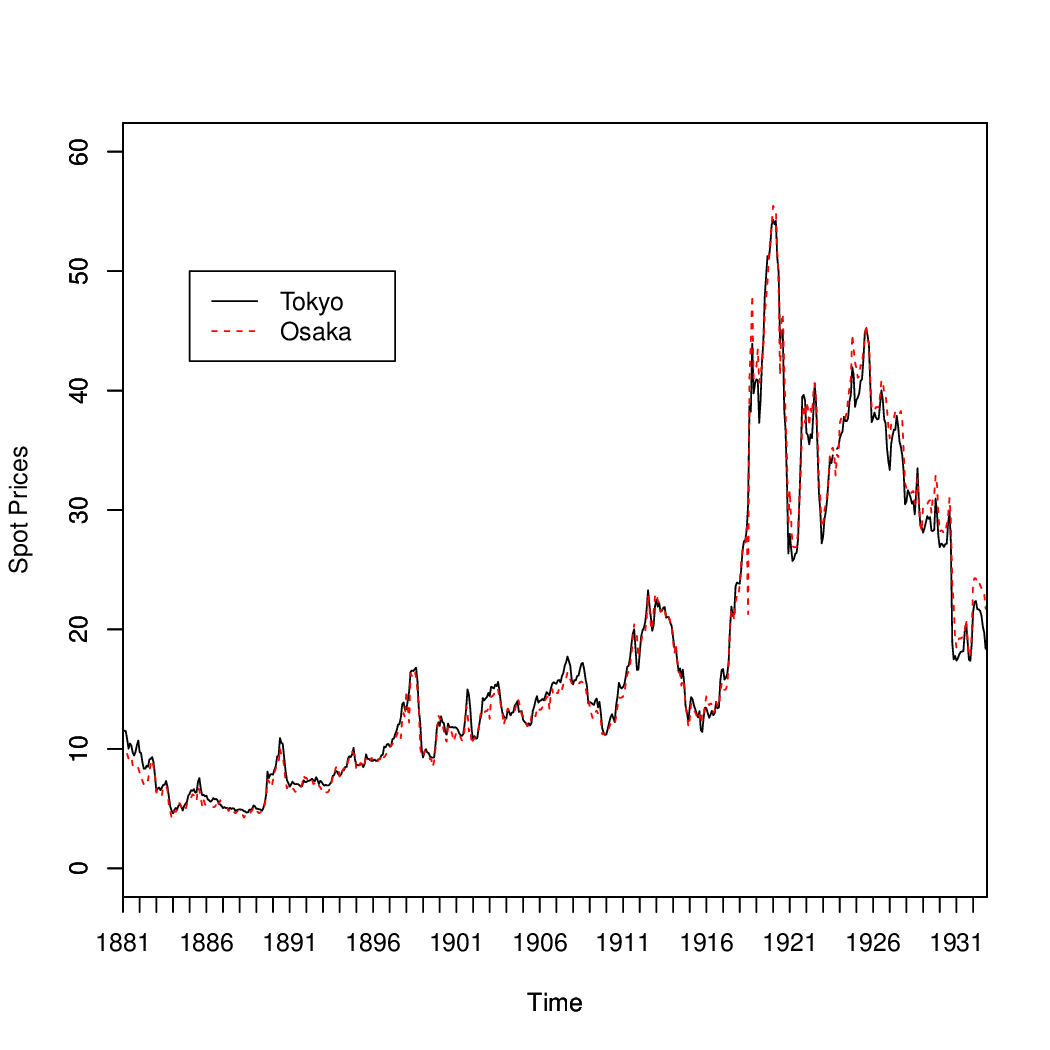}
\vspace*{3pt}
{\resizebox{12cm}{!}{
\begin{minipage}{325pt}
\footnotesize
\underline{Note}: R version 3.3.2 was used to compute the statistics.
\end{minipage}}}%
\end{center}
\end{figure}

Since the 1890s, Japan had been a continuous importer of rice. In the 1890s, Japan imported rice from countries in Southeast Asia, but from the 1900s, the amount of rice imported from Japanese colonies increased.\footnote{Japan officially colonized Taiwan in 1895 and Korea in 1910.} These colonies were the supply centers of rice to the Japanese home islands. In fact, Taiwan and Korea supplied a large portion of imported rice in Japan; colonial rice amounted to 48 per cent of total imported rice in the 1910s (see {\citet[pp.4--5]{maf1932drs}}). The Ministry of Agriculture and Commerce, which held jurisdiction over the administration of commodity exchanges, often ordered the exchanges to change transaction rules regarding the delivery of imported rice after 1890. These orders were aimed at suppressing the spot price of rice. Figure \ref{ei_fig1} shows changes in the spot price of rice in Tokyo and Osaka from 1881 to 1932.
\begin{center}
(Figure \ref{ei_fig1} around here)
\end{center}
We find that the spot price of rice jumped in 1890. On 17 April 1890, Masayoshi Matsukata, the finance minister, said, ``The rice price increases significantly more than the price of any other commodity, and people live in dire poverty. Alongside increases in rice prices, workers will force manufacturers to raise their wages. In addition, they maybe provoke riots or commit theft'' (see \citet[p.182]{ota1938rpp}).'' In summary, the government feared that the increase in rice prices would result in cost-push inflation and deterioration of the security situation. Accordingly, the Ministry of Agriculture and Commerce had to suppress rice prices. They forced the exchanges to accept imported rice as deliverable goods in order to address the increase in the supply of the deliverable goods in the rice exchanges. Because traders often delivered physical rice to clear their futures contracts on the exchanges, the Ministry of Agriculture and Commerce experimented with suppressing the spot price of rice to change the delivery rule in the futures markets. The government continued intermittently to force the exchanges to accept deliveries of imported rice. Table \ref{ei_table1} reports the periods in which the ministry issued such orders and the content of each order.
\begin{center}
(Table \ref{ei_table1} around here)
\end{center}
According to Table \ref{ei_table1}, we find that the Ministry of Agriculture and Commerce often ordered the exchanges to regard the delivery of imported rice. Whereas these orders were aimed at suppressing rice prices, the rice exchanges opposed these amendments. For example, in 1912, when the Ministry of Agriculture and Commerce forced the rice exchanges to steadily accept imported rice from Taiwan and Korea as an alternative to listed domestic rice, rice exchanges all over Japan released a statement denying any amendment to the transaction rule. Specifically, the exchanges said:

\begin{quote}
The rice futures price will fail to be an acceptable index of the expected spot price of rice because Taiwanese and Korean rice is different in quality from domestic rice.\footnote{See \citet[p.107]{maf1959haf}.}
\end{quote}

Taiwanese and Korean rice was the indica variety whereas domestic rice was the japonica variety. In addition, traders had other complaints about rice from Taiwan and Korea. Taiwanese rice varied considerably in quality because rice farming in Taiwan used double cropping (see \citet[p.27]{maf1938rtf}). Meanwhile, Korean rice was often intermixed with sand, stones, and other trash (see \citet[p.598]{hishimoto1938skr}). Korean rice was concentrated in Osaka because only Osaka's rice traders could decontaminate Korean rice (see \citet[p.20]{maf1938rtf}). Indeed, the imported rice was mainly consumed by poor people, such as coal miners and farmers in impoverished rural areas, as it was cheaper than domestic rice (see \citet[pp.59--64]{mochida1970drm} and \citet[pp.68--71]{omameuda1993fpm}). In 1912, the price per {\it koku} of domestic rice was 20.70 yen, while the price of Taiwanese (Taipei) rice was 16.77 yen and that of Korean (Busan) rice was 17.63 yen (see \citet[p.92]{maf1914rrr}). In summary, the traders and consumers did not regard the imported rice as being of the same quality as domestic rice.

In practice, although the Ministry of Agriculture and Commerce expected the amendment of the transaction rule to rein in rice prices, the futures price of rice did not fall in 1912. The Ministry of Agriculture and Commerce was puzzled by this situation (see \citet{asahi1912crp}). In 1918, it forced the exchanges to accept the delivery of low-quality domestic rice and to change standard rice from medium quality to low quality. The purpose of this change was also to control rice prices. In summary, despite these ordered changes to transaction rules, the government was unable to control rice futures prices, and the exchanges recognized that the only effect of the government intervention was to cause a dislocation in the rice market.

In 1918, Japan experienced nationwide riots, {\it kome-soudou} (``the Rice Riots''), which caused the Terauchi cabinet to resign and left Japan in disarray. New legislation was then required to manage this rice price inflation. The {\it Beikoku-hou} (``Rice Law''), was passed in 1921 and enabled the government to intervene directly in the spot markets to adjust prices. The policy was intensified in 1925 and 1931 through revisions to the law and was further modified by the the {\it Beikoku-tousei-hou} (``Rice Supply and Demand Regulation Law''), issued in 1933. Thus, the rice futures markets, which were supposed to be free, were increasingly controlled by the government from 1921.

\section{The Methods}\label{ei_sec3}

In this section, we present our empirical method for examining the market efficiency of rice futures markets in the Meiji, Taisho, and prewar-Showa periods in Japan. We consider these entities as established financial markets. Our method is based on  a non-Bayesian time-varying model, following \citet{ito2014ism}. The approach aims to examine market efficiency in the two rice futures markets, which were almost 250 miles (400 kilometers) apart and their market efficiency might have varied.

\citet{fama1970ecm} asserts that the financial commodity price in an efficient market instantly reflects any instantaneous shock. In other words, there is no persistent propagation of such a shock on the rate of return in an efficient market. Based on this contention, our analytical approach begins with impulse response analysis of rates of return in time-series data. Since the impulse response reflects the propagation of an instant shock to the system, the aggregation of the impulse responses represents the full effects of the shock.

It is standard to employ a VAR model with finite lags when attempting to obtain impulse responses for multivariate data:
\begin{equation}
{\ve y}_{t}={\ve\nu}+A_{1}{\ve y}_{t-1}+\cdots +A_{p}{\ve y}_{t-p}+{\ve u}_{t};\text{ \ }t=1,2,\ldots ,T, \label{VAR}
\end{equation}%
where $\ve{y}_t$ is a vector of rates of return for rice futures on the Tokyo and Osaka exchanges.

From equation (\ref{VAR}), a VMA($\infty$) model representing the impulse responses is obtained through a series of calculations. Note that we regard the VAR($p$) model as a reduced form of the VMA($\infty$).
\begin{equation}
{\ve y}_{t} ={\ve\mu} +\Phi _{0}{\ve u}_{t}+\Phi _{1}{\ve u}_{t-1}+\Phi _{2}{\ve u}_{t-2}+\cdots
;\text{ \ }t=1,2,\ldots ,T, \label{VMA}
\end{equation}
As mentioned earlier in this section, we calculate $\sum_{\tau=0}^\infty \Phi_t$ given VMA($\infty$) and then derive a norm from the matrix as a measure of the degree of persistent effects of the instantaneous shock. Following \citet{ito2014ism}, we adopt the spectral norm of the difference between $\sum_{\tau=0}^\infty \Phi_t$ and an identity matrix $I$, $\sum_{\tau=0}^\infty \Phi_t - I$.  Since $\Phi_0$ is identical to $I$, it reflects all effects except those of the shock.

In addition, our approach is drawn from the time-varying estimation technique of \citet{ito2014ism}. This technique allows us to obtain impulse responses at each period supposing a VAR($p$) model. Then we can derive impulse responses, $\{\Phi_{0,t}, \Phi_{1,t}, \Phi_{2,t}, \cdots\}_{t=1,\cdots,T}$, depending on time $t$.

\begin{equation}
{\ve y}_{t}={\ve\mu}_{t}+\Phi_{0,t}{\ve u}_{t}+\Phi_{1,t}{\ve u}_{t-1}+\Phi_{2,t}{\ve u}_{t-2}+\cdots \label{TVVMA}
\end{equation}%
Finally, we compute each degree of market efficiency $\zeta_t$ at $t=1,\cdots,T$ as follows.
\begin{equation}
\zeta_t=\sqrt{\max \ \lambda\left[(\sum_{\tau=0}^\infty \Phi_{\tau,t}-I)'(\sum_{\tau=0}^\infty\Phi_{\tau,t}-I)\right]}, \label{dist}
\end{equation}%
where ``$\max \ \lambda[X]$'' denotes the maximum eigenvalue of a matrix $X$. The sequence of $\zeta_t, (t=1,\cdots,T)$ provides information about the time-varying nature of the efficiency of the rice futures markets'. Since this method is based on a linear time-series model, a usual residual bootstrap technique can be used to facilitate statistical inference regarding the time-varying impulse responses and the degree of market efficiency.

\section{Data}\label{ei_sec4}

For this analysis, we use a dataset of monthly Tokyo and Osaka rice futures prices during and after the Meiji period, from January 1881 through November 1932. This is an old continuous monthly dataset for a well-established futures market.\footnote{The daily dataset for Tokyo and Osaka rice futures prices is not available before November 1911 (see \citet{maf1937drr}). Therefore, we construct a monthly dataset to analyze the time-varying structure of the futures markets using long-range data.} We employ monthly data for arguing the weak form of efficiency for the two major rice futures markets. We stress that it would be no problem to employ the data for our goal. The weak form of efficiency is mathematically based on a martingale series of a financial commodity's return. A martingale series is invariant under some wide range of transformations. For instance, when an original daily series is a martingale, the monthly series obtained by monthly averaging the daily series is still a martingale series. Furthermore, another monthly series obtained by selecting the last observation for each month is also still a martingale series. That is, employing lower frequency data due to availability would cause no problem except reducing precision for statistical analyses when we address the weak form of efficiency. In fact, many preceding studies on the relationship between commodity futures and economic policies employ monthly data (see, for instance, \citet{awokuse2003irc} and \citet{bhar2008icc}). 

\begin{figure}[bp]
 \caption{Futures Prices in Tokyo and Osaka Rice Markets}\label{ei_fig2}
 \begin{center}
 \includegraphics[scale=0.7]{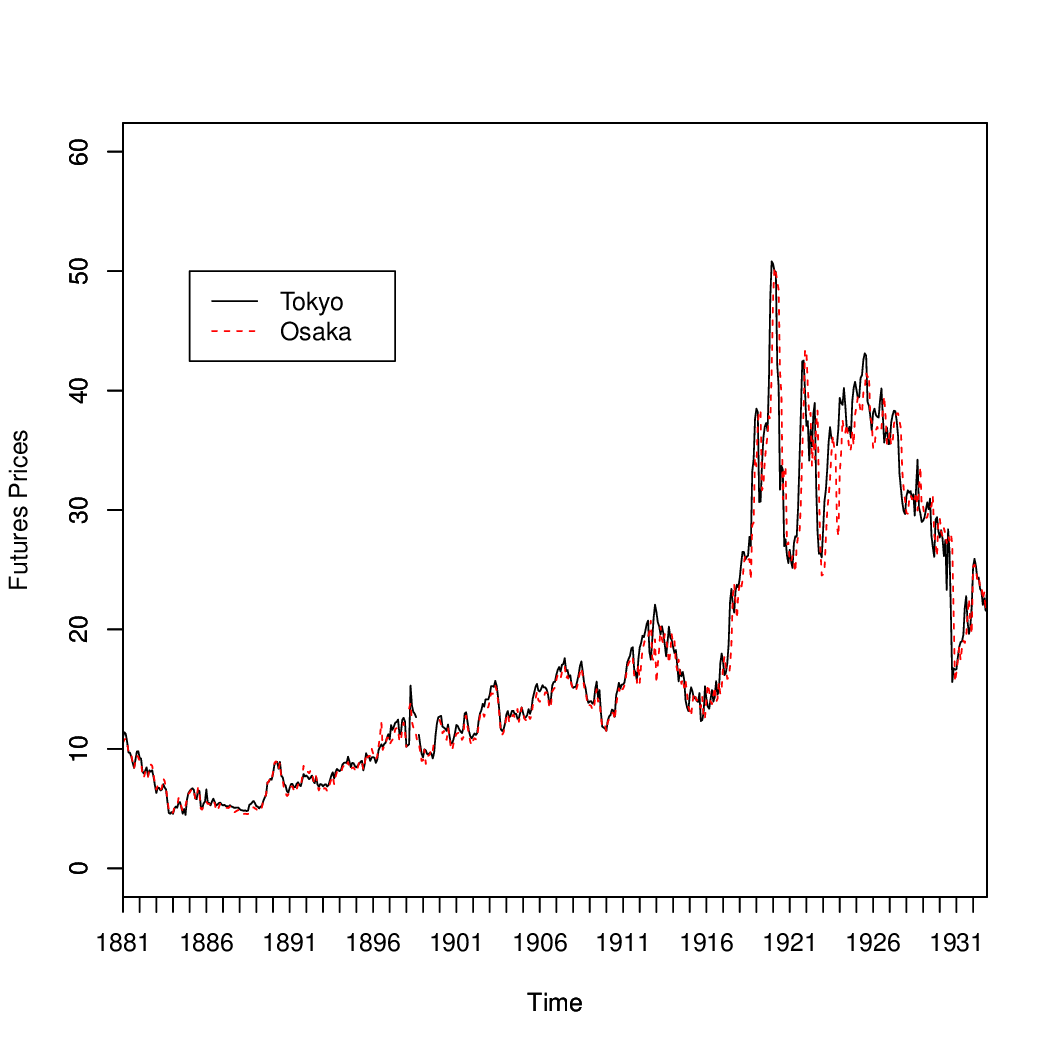}
\vspace*{3pt}
{\resizebox{12cm}{!}{
\begin{minipage}{325pt}
\footnotesize
\underline{Note}: R version 3.3.2 was used to compute the statistics.
\end{minipage}}}%
\end{center}
\end{figure}

For rice futures prices in Tokyo and Osaka, we utilize weighted average monthly values for the deferred contract (three months). The dataset is mostly that used in \citet{nakazawa1933nbh}, obtained mainly from the {\it Annual Statistical Report of the Tokyo Chamber Commerce} (Tokyo) and {\it Osaka City Statistics} (Osaka). There were a few missing values in the statistics created by \citet{nakazawa1933nbh}, some of which are sequential, with the longest missing sequences being for three months. These missing values are filled in using a seasonal Kalman filter, and we take the first difference of the natural log of rice futures prices to compute the ex-post return series. Figure \ref{ei_fig2} shows the variation of futures prices in the Tokyo and Osaka rice markets.
\begin{center}
(Figure \ref{ei_fig2} around here)
\end{center}
In the estimations, all variables appearing in the regression equations are assumed to be stationary. To check whether the variables satisfy the stationarity condition, we apply the augmented Dickey–Fuller generalized least squares (ADF-GLS) test of \citet{elliott1996eta}. Together with the procedure proposed by \citet{ng2001lls}, this unit root test is robust to size distortions. Table \ref{ei_table2} shows the results of the unit root test along with descriptive statistics for the data. The ADF-GLS test rejects the null hypothesis that the variable contains a unit root at the 1\% significance level.\footnote{For selecting the optimal lag length, we employ the modified Bayesian information criterion (MBIC) instead of the modified Akaike information criterion (MAIC) because we are unable to identify size distortions in the estimated coefficient of the detrended series, $\hat\psi$ (\citep{elliott1996eta} and \citet{ng2001lls}).}
\begin{center}
(Table \ref{ei_table2} around here)
\end{center}

\section{Empirical Results}\label{ei_sec5}

\subsection{Measuring the Time-Varying Degree of Market Efficiency}
Following the methods described in Section \ref{ei_sec3}, we first verify that the TV-VAR model is more appropriate than the traditional, time-invariant VAR model. In particular, we estimate the time-invariant VAR model with the whole sample and then apply \citetapos{hansen1992a} parameter constancy test to investigate whether the time-invariant model is a better fit for our data.
\begin{center}
(Table \ref{ei_table3} around here)
\end{center}
In order to select the lag lengths, we adopt the Bayesian information criterion (see \citet{schwarz1978edm}). Table \ref{ei_table3} presents the time-invariant VAR estimates and \citetapos{hansen1992a} joint parameter constancy test statistics (in the last row, ``$L_C$''). For our time-invariant VAR model, the parameter constancy test rejects the null hypothesis of constancy at the 1\% significance level against the alternative hypothesis that the parameter variation follows a random-walk process. These results suggest that the time-invariant VAR model does not apply to our data and that the TV-VAR model is a better fit.

Our measure of market efficiency captures how much the observed market deviates from an efficient market. Thus, we can clarify how the underlying market evolves by examining the degree of change over time. As mentioned in Section \ref{ei_sec3}, we focus on the TV-VAR model and the amount of market efficiency associated with this model. Once again, the deviation from efficient conditions in these futures markets is measured by (\ref{dist}), the spectral norm. If $\zeta_{t}=0$ for time $t$, the two futures markets are jointly efficient.

\begin{figure}[bp]
 \caption{The Time-Varying Degree of Market Efficiency}\label{ei_fig3}
 \begin{center}
 \includegraphics[scale=0.7]{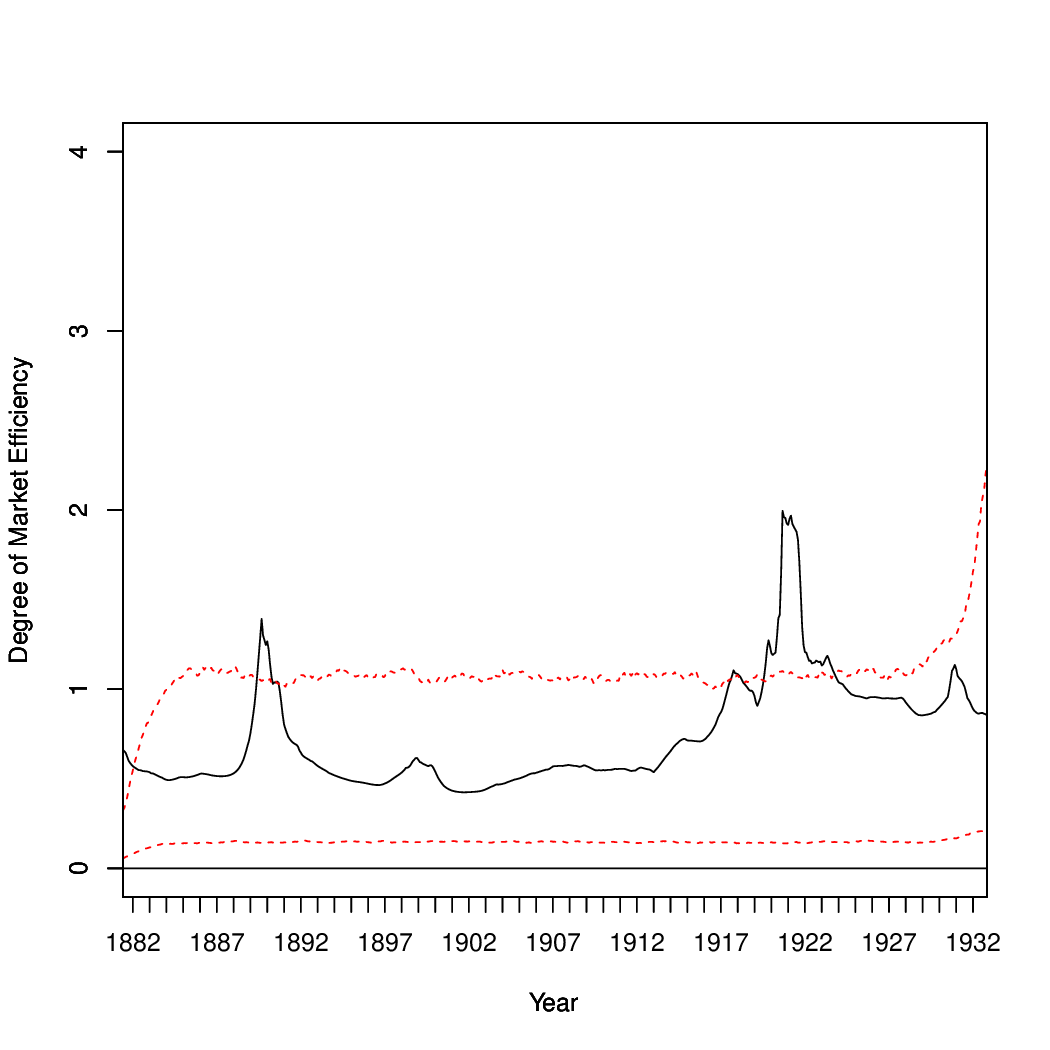}
\vspace*{3pt}
{\resizebox{12cm}{!}{
\begin{minipage}{325pt}
\footnotesize
\underline{Notes}:
\begin{itemize}
\item[(1)] The dashed red lines represent the 95\% confidence bands of the 
           time-varying spectral norm in the case of an efficient market. 
\item[(2)] We ran bootstrap sampling 5000 times to calculate the confidence 
           bands.
\item[(3)] R version 3.3.2 was used to compute the estimates.
\end{itemize}
\end{minipage}}}%
\end{center}
\end{figure}

We compute the degree of market efficiency from the estimates of $\Phi_{t}\left(1\right)$; therefore, $\zeta_{t}$ is subject to sampling error. Thus, we include a confidence band for $\zeta_{t}$ under the null hypothesis of market efficiency. We find no evidence of inefficient markets whenever the estimated $\zeta_{t}$ is less than the upper limit of the confidence band; inefficient markets are detected with an estimated $\zeta_{t}$ that is larger than the upper limit. Following \citet{ito2014ism}, we adopt a residual-based bootstrapping technique for our TV-VAR model to construct the confidence interval. This consists of three steps: (i) we generate multivariate independent and identically distributed processes for futures returns, $y_{t}$; (ii) we apply and estimate the TV-VAR model for those processes; and finally, (iii) we compute $\zeta_{t}$.
\begin{center}
(Figure \ref{ei_fig3} around here)
\end{center}
Figure \ref{ei_fig3} shows that the examined rice futures markets are generally efficient. However, it also shows that the degree of market efficiency varies over time: it is relatively lower starting in the late 1910s, and the markets were inefficient around 1890 and in the early 1920s. These changes in market efficiency stem from government interventions in the exchanges' transaction rules for rice futures.

\section{Historical Interpretation}\label{ei_sec6}

\subsection{The Amendment of Transaction Rule in 1890}
In 1890, an amendment to the transaction rule forced the rice exchanges to accept the delivery of imported rice for the first time (see Table \ref{ei_table1}). The government imposed the amendment of the rule on the exchanges in order to address the increase in the supply of the deliverable goods, as mentioned in Section \ref{ei_sec2}. In short, this intervention was a simple policy regarding rice supply control. However, it reduced the efficiency of transactions on the futures markets of Tokyo and Osaka, and their traders showed strong opposition to the amendment. In May 1890, the Tokyo Rice Wholesalers Association issued a statement of strong disagreement with the amendment to the transaction rule, as follows.

\begin{quote}
``Imported rice is different from domestic rice in quality and use. If the exchanges accept the delivery of imported rice, movements of futures prices will not be similar to those in the spot prices of domestic rice. Therefore, the futures price will fail to be a fine index of the expected price of rice. We expect that the rice exchange will become instead a gambling place.''\footnote{See \citet[p.577]{tokyo1890roa}).}
\end{quote}

As mentioned in Section \ref{ei_sec2}, because the imported rice was different in quality from domestic rice, Japanese consumers did not regard imported and domestic rice the same. The distribution of imported rice within Japan differed between regions, with more found in Osaka than in Tokyo. From October 1889 to June 1890, the volume of rice imported into the ports of Osaka and nearby Kobe was 482,000 {\it koku} in total, while that for Yokohama, near Tokyo, was 247,000 {\it koku}. The volume of rice imported into Osaka and Kobe accounted for about 68 per cent of the total volume of rice imported in the same period (see \citet{tokyo1890sri}). Therefore, the increase in rice imports had a greater impact on the rice market of Osaka than on that of Tokyo. The Osaka-Dojima rice exchange opposed changes to trading rules that allowed imported rice to be deliverable as an alternative listed good. The Ministry of Agriculture and Commerce revoked the exchange's permit to deal in reaction to its disobedience (see \citet{tokyo1890rpd}). Dealing in the Osaka-Dojima rice exchange stopped temporarily, and this series of serious disruptions reduced market efficiency still further. While traders and consumers did not regard the imported rice as being of the same quality as domestic rice, buyers in the exchanges were liable to obtain imported rice whose quality was lower than the standard rice after the amendment of the transaction rule. The buyers could not choose the delivery rice, and they faced uncertainty about the rice delivery in the exchanges. This situation forced the traders to face great uncertainty, since the sellers often delivered physical rice to buyers when they cleared their transactions in futures, as mentioned in Section \ref{ei_sec2}. In fact, the Ministry of Finance reported that the amendment to the transaction rule suppressed futures prices, although the spot price did not change (see \citet[p.194]{mof1919hrp}). In other words, the futures price failed to be a fine index of the expected price of rice, as the Tokyo Rice Wholesalers' Association had predicted. When imported rice was temporarily made deliverable for a second time in 1898, the intervention caused a smaller disruption to the exchanges than it had in 1890. Thus, while the exchanges experienced low market efficiency, there was no significant change in the degree of market efficiency, according to our analysis.

\subsection{The Amendment of Transaction Rule in 1898}
In 1898, rice prices soared because Japan took a hit from a poor rice harvest in the previous year. In January 1898, the Ministry of Agriculture and Commerce ordered the exchanges to change the transaction rule regarding the delivery of imported rice as a way of suppressing the price increases. However, this amendment of the transaction rule had poor efficacy in terms of price control due to a record harvest of rice and barley in 1898. In prewar Japan, barley was widely used as a substitute for rice and thus, also impacted rice prices. The two crops had different seasons: rice was grown in fall, while the harvest season for barley was grown in early summer. In May 1898, newspapers reported that the barley harvest would be a good one (see \citet{asahi1898fbc}). In July 1898, the newspapers reported a forecast for a good rice harvest (see \citet{yomiuri1898rpf}); indeed, the Ministry of Agriculture and Commerce reported the resulting harvest as the biggest crop since record-keeping began (See \citet[p.1]{mac1900sya}).

The total volume of rice produced in 1898 was 4.7 million {\it koku}, 43\% higher than in the previous year (see \citet[pp.1--2]{mac1900sya}). Under this supply situation, rice traders began to perceive excessive rice imports starting in May 1898, when the volume of imported rice inventory increased in the warehouses at the port of Yokohama (see \citet{yomiuri1898sir}). Moreover, the newspapers reported forecasts of good harvests for domestic rice and barley. The market environment disrupted transactions in futures. Thus, the exchanges failed to offer a fine index of the expected price of rice. In August 1898, the {\it Yomiuri Shimbun} reported this situation as follows.

\begin{quote}
``After January, when the delivery of imported rice in the rice exchanges began, the rice futures market was dependent on price fluctuations for imported rice. As a result, the trend in rice futures prices was different from that in the spot price.''\footnote{See \citet{yomiuri1898cad}.}
\end{quote}

On the other hand, rice traders, represented by the Tokyo Rice Wholesalers Association, did not issue a statement of strong disagreement with the transaction rule amendment, as was the case in 1890. Because imported rice was not being traded actively, government intervention caused a smaller disruption in the exchanges in 1898 than it had in 1890.

\subsection{Frequent Government Interventions in the 1910s}
The Ministry of Agriculture and Commerce very often ordered the exchanges to change transaction rules related to the delivery of Taiwanese and Korean rice starting in 1912. In 1913, this imported rice began to be deliverable on a steady basis; the amendment related to rice from Taiwan was abolished, while that for rice from Korea was retained. In addition, lower quality domestic rice became deliverable after the Ministry of Agriculture and Commerce extended the variety of rice deliverable in the rice futures exchanges in 1918 and 1919. Imported rice from Taiwan became deliverable again in 1919 as part of a government intervention to suppress rice prices (see Table \ref{ei_table1}). This series of government interventions in rice futures exchanges reduced their market efficiency. The rice traders opposed this series of government interventions, as mentioned in Section \ref{ei_sec2}. In addition, some government officials admitted the failure of their interventions in rice futures market. In 1918, Yoshinari Kawai, Director of the Division of Foreign Rice Management at the Ministry of Agriculture and Commerce, who was responsible for controlling the rice market to stabilize prices and transactions, said that the rice futures market faced deterioration:

\begin{quote}
``Essentially, the price difference between the spot and the futures prices vanishes at the maturity date. However, after the exchanges began to regard the delivery of imported rice, the price difference between the spot and futures prices did not fall with the maturity date. The futures market showed hardly any relationship with the spot market because the rice futures market was strongly dependent on the price of rice from Taiwan and Korea. The function of the exchange is thus disrupted.''\footnote{See \citet[pp.300--301]{kawai1921loe}.}
\end{quote}

Thus, regarding the policy summarized above, Kawai conceded that it was not functional. In addition, between 1918 and 1921 the Ministry of Agriculture and Commerce often forced the exchanges to extend the variety of rice deliverable and to change the standard rice in the rice futures exchanges; as a result, the market efficiency fell markedly. In fact, the Tokyo Asahi Shimbun reported that these changes to the transaction rules greatly enlarged the price difference between the spot and futures prices after 1918 (see \citet{asahi1920rrp}). In other words, a reduction in market efficiency occurred in the 1910s and early 1920s because the Ministry of Agriculture and Commerce forced the exchanges to amend the transaction rules frequently. This was especially true from 1918 to 1921, when the Ministry of Agriculture and Commerce intervened in the rice futures exchanges frequently.

The Rice Law, based on lessons and reflections from the Rice Riots in 1918, was established in 1921. The Rice Law permitted the government to buy and sell physical rice directly to adjust the supply-demand balance as well as rice prices. When the Great Kanto Earthquake occurred in Tokyo in September 1923, the Japanese government announced the purchase of 1.35 million {\it koku}, or 2.45\% of domestic production in 1922 of 55.18 million {\it koku}, to be sold to earthquake victims. While the amount actually purchased was 0.59 million {\it koku}, the government strongly affected prices in the rice exchanges by announcing the direct purchase of such a large volume (see \citet[p.164]{maf1959haf}).

About 100,000 people in Tokyo and surrounding areas died in the Great Kanto Earthquake in 1923. In addition to this massive human suffering, the Tokyo rice exchange building was destroyed, and rice futures dealings stopped for two months. In addition, a fire destroyed 75\% of the government's rice, or 267,000 {\it koku}, which had been stored in the area surrounding Tokyo. Therefore, the government transported 456,000 {\it koku} of rice from western Japan to help the earthquake victims and sold it at the official price in Tokyo, Yokohama, and Yokosuka. Because of these disruptions to the rice market, market efficiency of the rice exchanges fell markedly. In particular, the government attempted to control the supply–demand of the rice spot market in the aftermath of earthquake. Under this situation, traders in the exchanges could no longer expect fluctuations in the rice price, since the policy to help the earthquake victims affected the price formation of rice. That is, traders faced great uncertainty in the exchanges. However, after the exchanges restarted in November 1923, their market efficiency again improved despite the delivery of Korean rice on the exchanges (see Table \ref{ei_table1}), resulting from differences in quality between domestic and Korean rice.

\subsection{The 1920s Policy for Uniform Rice Quality}
The Governor-General of Korea promoted japonica rice cultivation in the colony beginning in the 1910s. Japonica production increased from only 5\% of total rice production in 1912 to 69\% in 1921 and 79\% in 1932 (see \citet[pp.438--439]{tobata1939rek}). This policy resulted in shrinking the difference between rice prices in Korea and those in inland Japan from the 1920s, as Japanese consumers began to regard rice imported from Korea as being of the same quality as domestic rice (see \citet[p.195]{omameuda1993fpm}). As a result, even when rice imported from Korea was delivered for futures transactions, market participants were no longer worried about the risk of delivery of low-quality rice. In fact, the exchanges recognized that the quality of Korean rice had improved. From January 1927, all rice exchanges in Japan raised their grade of Korean rice (see \citet{asahi1926grd}). After the Great Kanto Earthquake, the market efficiency of the two major rice futures markets improved.\footnote{Note that market efficiency was relatively low during the post-disaster reconstruction period because the government intermittently bought and sold rice under the Rice Law.}

\section{Conclusion}\label{ei_sec7}

This study measured the time-varying joint degree of market efficiency of the two major rice futures markets in prewar Japan and investigated the factors that altered market efficiency. It argued that Japan's two major rice futures markets were nearly efficient, although there were several periods during which government interventions concerning trading rules reduced their market efficiency.

Starting in 1890, in order to suppress rice prices, the government forced the rice exchanges to accept deliverable imported rice as an alternative to domestic rice. However, before the early 1920s, rice imported from Japanese colonies was of notably different quality to domestic rice. Thus, most rice traders did not regard this rice as being of the same quality as standard domestic rice. Ignoring the context of rice futures dealings, the government forced the exchanges to regard imported rice from Taiwan and Korea as deliverable rice on the futures exchanges, resulting in reduced efficiency in rice futures dealings for several periods. When rice prices rose quickly in the late 1910s, the government extended the range of deliverable rice on the rice futures exchanges to reduce the price of physical rice. Since newspapers at the time reported that the intervention expanded the gap between spot and future prices, the market efficiency of rice futures dealing fell. Thus, frequent government interventions in the markets reduced the efficiency of rice futures during the 1910s. However, from the mid-1920s, the rice futures markets improved their efficiency due to the decrease in the quality difference between domestic rice and imported rice from Korea, after the Governor-General of Korea promoted domestic rice cultivation.

From the 19th century, the demand for food increased rapidly in Japan alongside the country's industrialization and urbanization. Thus, an important task for the government was to procure food. In particular, the Japanese government accelerated food imports from the country’s colonies and promoted the distribution of imported food in the home islands of Japan. However, there was a difference in quality between imported food from the Japanese colonies and domestic food. Consequently, when the government forced the exchanges to regard colonial rice as being of the same quality as domestic rice, the intervention disrupted the rice markets in Japan. The government interventions, which promoted domestic distribution of colonial goods, resulted in confusion in the commodity markets, and decreased the efficiency of markets in the metropole.

\section*{Acknowledgments}

We would like to thank the editor, Stefano Battilossi, three anonymous referees, Shigehiko Ioku, Junsoo Lee, Kris Mitchener, Chiaki Moriguchi, Tetsuji Okazaki, Minoru Omameuda, Rainer Sch\"ussler, Masato Shizume, Yasuo Takatsuki, Tatsuma Wada, Wako Watanabe, and Asobu Yanagisawa for their helpful comments and suggestions. We would also like to thank seminar and conference participants at Wakayama University; the Japanese Economics Association 2014 Spring Meeting; the 84th Annual Conference of the Socio-Economic History Society; and the 89th Annual Conference of the Western Economic Association International for helpful discussions. We also thank the Japan Society for the Promotion of Science for their financial assistance, as provided through the Grant in Aid for Scientific Research Nos. 26380397 (Mikio Ito), 26780199 (Kiyotaka Maeda), and 15K03542 (Akihiko Noda). All data and analysis codes used for this study are available on request.


\setcounter{table}{0}
\renewcommand{\thetable}{\arabic{table}}

\clearpage

\begin{table}[tbp]
\caption{Government Interventions in the Rice Futures Exchanges}\label{ei_table1}
\begin{center}
  \begin{tabular}{cllp{110mm}c}\hline\hline
   & Date & & Orders and Amendments & \\\hline
   & May, 1890 & & Rice futures exchanges accept imported rice as an alternative to listed domestic rice. & \\
   & November, 1890 & & The amendment of May 1890 is abolished. & \\
   & January, 1898 & & The amendment of May 1890 is revived. & \\
   & October, 1898 & & The amendment of January 1898 is abolished. & \\
   & June, 1912 & & Rice futures exchanges accept imported rice from Taiwan and Korea as an alternative to listed domestic rice. & \\
   & October, 1912 & & The amendment of June 1912 is abolished. & \\
   & March, 1913 & & The exchanges accept imported rice from Taiwan and Korea as an alternative to listed domestic rice on a steady basis. & \\
   & August, 1914 & & Rice from Taiwan is undeliverable. & \\
   & April, 1918 & & Low-quality domestic rice is deliverable and the exchanges accept a change to the standard rice from medium quality to low quality. & \\
   & February, 1919 & & Rice from overseas is deliverable. & \\
   & November, 1919 & & Lowest quality domestic rice is deliverable and the exchanges accept a change to standard rice from low quality to lowest quality. & \\
   & December, 1919 & & Rice from overseas (excluding Korea) is undeliverable. & \\
   & October, 1920 & & The exchanges accept a change to standard rice from lowest quality to medium quality. & \\
   & December, 1920 & & Lowest and low-quality domestic rice is undeliverable. & \\
   & November, 1921 & & Low-quality domestic rice is deliverable. & \\\hline\hline
\end{tabular}
\vspace*{5pt}
{\resizebox{15cm}{!}{
\begin{minipage}{550pt}
{\underline{Sources:}}
 \begin{itemize}
 \item[(1)] \citet{boj1957hmh}, ``1899 Statistical Yearbook of Bank of Japan,'' in Nihon Kinyuushi Shiryo Meiji-Taisho Hen Dai 19 Kan [Materials on Japanese Financial History in the Meiji-Taisho Period Vol.19], Printing Bureau of Ministry of Finance, Tokyo, Japan.
\item[(2)] \citet{mof1919hrp}, ``History of Rice Price Adjustment in the Meiji-Period,'' Tokyo, Japan.
\item[(3)] \citet{maf1959haf}, ``History of Agricultural and Forest Administration, Vol. 4,'' Tokyo, Japan.
 \end{itemize}
\end{minipage}}}%
\end{center}
\end{table}

\clearpage

\clearpage

\begin{table}[tbp]
\caption{Descriptive Statistics and Unit Root Tests}\label{ei_table2}
\begin{center}\resizebox{15cm}{!}{
\begin{tabular}{cccccccccccccc}\hline\hline
& & & Mean & SD & Max & Min & & ADF-GLS & Lags & $\phi$ & & $\mathcal{N}$ & \\\cline{4-7}\cline{9-11}\cline{13-13}
& Tokyo & & 0.0012 & 0.0626 & -0.3189 & 0.3860 & & -20.4604 & 0 & 0.1930 & & 622 &  \\
& Osaka & & 0.0013 & 0.0633 & -0.3069 & 0.2523 & & -20.6532 & 0 & 0.1841 & & 622 &  \\\hline\hline
\end{tabular}}
\vspace*{5pt}
{\resizebox{14cm}{!}{
\begin{minipage}{550pt}
{\underline{Notes:}}
\begin{itemize}
\item[(1)] ``ADF-GLS'' denotes the ADF-GLS test statistics, ``Lags'' 
           denotes the lag order selected by the MBIC, and ``$\hat\phi$'' 
           denotes the coefficients vector in the GLS detrended series 
           (see Equation (6) in \citet{ng2001lls}).
\item[(2)] In computing the ADF-GLS test, a model with a time trend 
           and a constant is assumed. The critical value at the 1\% 
           significance level for the ADF-GLS test is ``$-3.42$.''
\item[(3)] ``$\mathcal{N}$'' denotes the number of observations.
\item[(4)] R version 3.3.2 was used to compute the statistics.
\end{itemize}
\end{minipage}}}%
\end{center}
\end{table}

\clearpage

\begin{table}[tbp]
\caption{Time-Invariant VAR Estimations}\label{ei_table3}
\begin{center}\resizebox{8cm}{!}{
\begin{tabular}{lllccc}\hline\hline
 &  &  & Tokyo & Osaka & \\\cline{4-5}
 & \multirow{2}*{Constant} &  & 0.0014 & 0.0010 & \\
 &  &  & [0.0024] & [0.0018] &  \\
 & \multirow{2}*{$R_{t-1}^{tokyo}$} &  & 0.0629 & 0.1170 & \\
 &  &  & [0.0720] & [0.0362] & \\
 & \multirow{2}*{$R_{t-1}^{osaka}$} &  & 0.1561 & 0.0385 & \\
 &  &  & [0.0701] & [0.0553] & \\
 & \multirow{2}*{$R_{t-2}^{tokyo}$} &  & -0.0268 & 0.5453 & \\
 &  &  & [0.0421] & [0.0792] & \\
 & \multirow{2}*{$R_{t-2}^{osaka}$} &  & -0.0263 & -0.4130 & \\
 &  &  & [0.0585] & [0.0517] & \\
 & \multirow{2}*{$R_{t-3}^{tokyo}$} &  & -0.1750 & 0.0495 & \\
 &  &  & [0.0599] & [0.0496] & \\
 & \multirow{2}*{$R_{t-3}^{osaka}$} &  & 0.0122  & -0.0916 & \\
 &  &  & [0.0472] & [0.0394] & \\
 & \multirow{2}*{$R_{t-4}^{tokyo}$} &  & 0.0300 & 0.2329 & \\
 &  &  & [0.0544] & [0.0606] & \\
 & \multirow{2}*{$R_{t-4}^{osaka}$} &  & -0.0174 & -0.0871 & \\
 &  &  & [0.0399] & [0.0393] & \\\hline
 & ${\bar R}^2$ &  & 0.0336 & 0.2644 & \\
 & $L_C$ &  & \multicolumn{2}{c}{68.7284} &  \\\hline\hline
\end{tabular}}\\
\vspace*{5pt}
{\resizebox{10cm}{!}{
\begin{minipage}{350pt}
{\underline{Notes:}}
 \begin{itemize}
\item[(1)] ``${\bar{R}}^2$'' denotes the adjusted $R^2$, and ``$L_C$'' denotes \citetapos{hansen1992a} joint $L$ statistic with variance.
\item[(2)] \citetapos{newey1987sps} robust standard errors are in brackets.
\item[(3)] R version 3.3.2 was used to compute the estimates and the test statistics.
 \end{itemize}
\end{minipage}}}%
\end{center}
\end{table}

\clearpage

\end{document}